\documentclass[letterpaper]{article} 
\usepackage[preprint]{aaai2027}  
\usepackage[hyphens]{url}  
\usepackage{graphicx} 
\usepackage{natbib}  
\usepackage{caption} 
\usepackage{algorithm}
\usepackage{algorithmic}

\usepackage{newfloat}
\usepackage{listings}
\DeclareCaptionStyle{ruled}{labelfont=normalfont,labelsep=colon,strut=off} 
\floatstyle{ruled}
\newfloat{listing}{tb}{lst}{}
\floatname{listing}{Listing}

\usepackage{booktabs}

\usepackage{amsfonts}
\usepackage{amsmath}
\usepackage{multirow}      
\usepackage{colortbl}      

\title{ViSSFlow: Unifying Video-Conditioned Sound and Speech Generation \\via Joint Learning}
\author{
}
\affiliations{
}
\author{
Xin Cheng\textsuperscript{\rm 1}\equalcontrib\thanks{Contact  chengxin000@ruc.edu.cn},
Yuyue Wang\textsuperscript{\rm 1}\equalcontrib,
Xihua Wang\textsuperscript{\rm 1}\equalcontrib,
Yihan Wu\textsuperscript{\rm 1},
Kaisi Guan\textsuperscript{\rm 1},
Yijing Chen\textsuperscript{\rm 1},
Peng Zhang\textsuperscript{\rm 2},
Kieran Liu\textsuperscript{\rm 2},
Meng Cao\textsuperscript{\rm 2},
Ruihua Song\textsuperscript{\rm 1}
}
\affiliations{
\textsuperscript{\rm 1}Renmin University of China
\textsuperscript{\rm 2}Apple\\
}

\begin{document}

\maketitle

\begin{abstract}
Video-conditioned audio generation, including Video-to-Sound (V2S) and Visual Text-to-Speech (VisualTTS), has traditionally been studied as two separate tasks despite their shared objective of generating audio conditioned on visual content. Existing methods rely on task-specific architectures and training paradigms, while the possibility of a unified generative framework remains largely unexplored due to heterogeneous conditioning signals and limited joint training data. In this paper, we propose ViSSFlow, a unified flow-matching framework for both V2S and VisualTTS. To effectively integrate diverse conditioning modalities within a Diffusion Transformer (DiT), we introduce a disentangled condition aggregation mechanism that exploits the complementary roles of attention layers, using cross-attention to model semantic conditions and self-attention to capture temporally dense visual cues. Furthermore, contrary to the common assumption that joint optimization of V2S and VisualTTS causes task interference, we show that a single model can be trained end-to-end without sacrificing task-specific performance. To address the scarcity of paired training data covering both sound and speech generation, we further propose a simple yet effective feature-level data synthesis strategy that enables robust joint learning. Extensive experiments on V2S, VisualTTS, and joint generation benchmarks demonstrate that ViSSFlow consistently outperforms state-of-the-art task-specific baselines while providing a unified solution for video-conditioned audio generation. Code and demos are provided in Supplementary Material.
\end{abstract}

\begin{figure*}[t!]
    \centering
    \includegraphics[width=.98\textwidth]{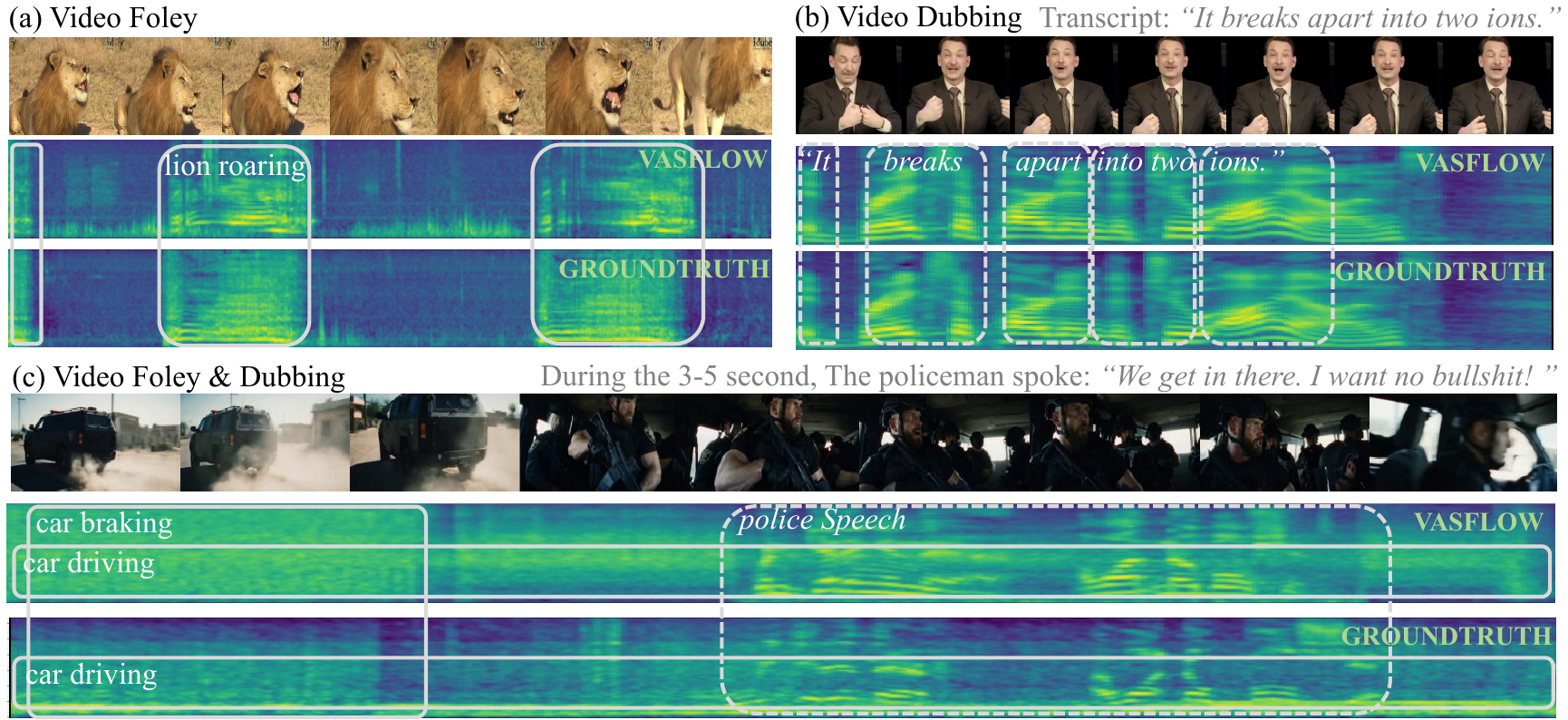}
    \caption{
    ViSSFlow: A unified generative model for video-conditioned sound and speech synthesis.
    (a) Video-to-Sound generation given a silent video.
    (b) Visual-TTS generation given a silent talking video and speech transcripts.
    (c) Sound-Speech Joint generation given a silent video and transcripts.
    }
    \label{fig:teaser}
\end{figure*}


\section{Introduction}
\label{sec:intro}

The rapid advances in generative models have made sound and speech synthesis fundamental components of immersive multimodal content creation~\cite{deepmind2024generating, googleveo32025google}. Among these tasks, video-conditioned audio generation, encompassing both Video-to-Sound (V2S~\footnote{In prior works, the term ``Video-to-Audio (V2A)" or ``Foley'' has commonly been used. To avoid ambiguity here, we define ``sound" as non-linguistic audio, such as environmental or natural audio, distinct from speech. ``Speech" refers to audio containing linguistic information, such as spoken language. ``Audio" is used to encompass sound, speech, and all other types of auditory content.})~\cite{cheng2024lova, cheng2025MMAudio} and Visual Text-to-Speech (VisualTTS)~\cite{cong2024styledubber, cong2024emodubber},  has recently attracted growing attention.
Although both tasks aim to generate audio (sound and speech, respectively) synchronized with visual input, they have largely been studied independently because they rely on different conditioning signals and target distinct generation objectives. Consequently, existing V2S models cannot synthesize intelligible speech, while VisualTTS models are unable to generate non-speech environmental sounds.

Several attempts have been made recently to unify video-conditioned sound, speech, and joint generation tasks~\cite{zhang2025deepaudio,tian2025dualdub, wang2025audiogenomni}. 
However, they either decouple sound and speech generation by employing separate modules, rely on complex curriculum learning procedures to acquire the two capabilities independently, or are unable to jointly generate speech and sound because of the scarcity of high-quality joint data. 
Consequently, developing an end-to-end framework unifying these tasks remains an open question, as the model architecture, training strategies, and data scarcity remain largely unaddressed.

To handle these challenges, we introduce ViSSFlow, a flow-matching~\cite{lipman2022flow}  framework  that unifies V2S, VisualTTS and video-conditioned sound-speech joint generation. 
First, we implement a disentangled condition aggregation mechanism to handle multiple heterogeneous condition sequences. 
We systematically explore the two typical types of conditioning mechanisms within the attention-based Diffusion Transformer (DiT)~\cite{Peebles2022DiT}: through cross-attention and self-attention. In ViSSFlow, we integrate temporally dense conditions (e.g., text transcript, sound and speech synchronization features) by concatenating them with audio latents and processing them through self-attention. By contrast, semantically rich video features are incorporated via cross-attention layers. We demonstrate the effectiveness of our conditioning mechanisms through extensive ablations.
Second, we observe that end-to-end full-parameter joint learning of sound and speech does not lead to performance degradation, a common issue observed in previous methods~\cite{kendall2018multi, tian2025dualdub}. ViSSFlow requires no complex design of separate modules or training stages, simplifies the pipeline and reduces complexity. 
Third, to address the scarcity of high-quality joint video-sound-speech data, we apply a simple yet effective feature-level data synthesis method, which mitigates data scarcity and improves the generation quality of both sound and speech.
Finetuned with the synthetic data, ViSSFlow can be easily adapted to real-world scenarios, generating sound and speech jointly (i.e., speech with environmental sounds) as shown in Fig.~\ref{fig:teaser}(c). Finally, evaluation on V2S, VisualTTS, and joint generation benchmarks shows that ViSSFlow achieves exceptional sound fidelity and speech quality, compared to domain-specific and pipeline-based methods. 

In summary, our contributions are as follows:
(1) We introduce ViSSFlow, a unified flow-based framework for video-conditioned sound and speech generation, with an effective condition aggregation mechanism to integrate multiple features for sound and speech generation into DiT blocks.
(2) We demonstrate that end-to-end full-parameter joint learning of sound and speech is feasible within ViSSFlow, eliminating the need for separate modules or complex multi-stage training strategies.
(3) We further show ViSSFlow can be easily adapted to joint sound-and-speech generation through a simple feature-level on-the-fly data synthesis strategy, which alleviates data scarcity and improves joint generation quality.


\section{Related Work}
\label{sec:related}

\begin{figure*}[t!]
    \centering
    \includegraphics[width=1.0\linewidth]{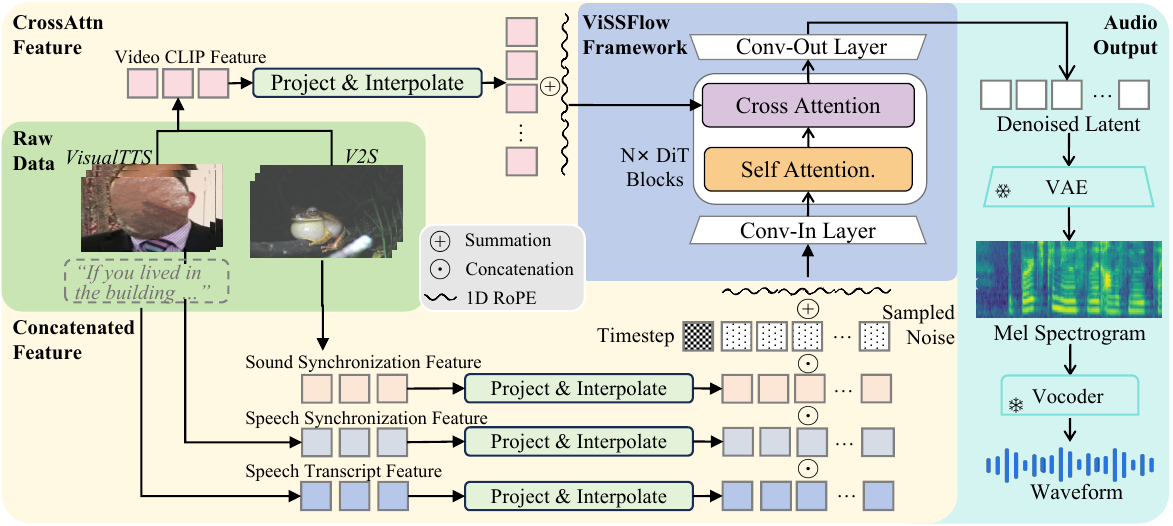}
    \caption{Architecture of ViSSFlow.
    ViSSFlow employs cross-attention-based DiT blocks and a flow-matching paradigm, which can handle multiple conditional inputs via cross-attention or concatenation with latents. Temporally dense signals like sound synchronization, speech synchronization, and speech transcript features are introduced by concatenation, while video semantic features are incorporated through cross-attention. 
    }
    \label{fig:arch}
\end{figure*}

\subsection{Video-to-Sound and Visual Text-to-Speech Generation}
Video-to-Sound (V2S) aims to generate semantically and temporally aligned environmental or natural sounds from video input. Recent models have four main paradigms: autoregressive~\cite{sheffer2022im2wav, Iashin2021Spec, mei2024foleygen, viertola2025vaura}, mask~\cite{liu2024tellvatt, pascual2024masked, Su2024beyond}, flow and diffusion~\cite{luo2023difffoley, zhang2024foleycrafter, wang2024tiva, cheng2024lova, wang2024frieren, xing2024seeing, cheng2025MMAudio}.
Most V2S methods rely on CLIP features~\cite{radford2021CLIP} for semantics and incorporate auxiliary temporal cues such as CAVP~\cite{luo2023difffoley}, SynchFormer~\cite{iashin2024synchformer}, and energy curves~\cite{jeong2025read} to improve alignment.
Another task, Visual Text-to-Speech (VisualTTS), focuses on generating speech that matches the speaker’s style, lip, emotion, and other visual cues. These models~\cite{chen2022v2c, cong2023hpmdubbing, cong2025emodubber, lu23fDSU, cong2024styledubber, hu2021neural} are typically built on strong TTS backbones~\cite{chien2021FastSpeech2, mehta2024matcha} and leverage multiple prosodic controls (reference speech, pitch, energy, emotion, etc.).
Despite their connection, V2S and VisualTTS are usually studied separately: V2S lacks intelligible speech generation capability, while VisualTTS cannot generate non-speech sound. Therefore, our work tries to unify the two generation tasks.

\subsection{Unified Video-Conditioned Sound and Speech Generation}
Some advancements have been made to integrate video-conditioned sound and speech generation into unified models.
Systems like Meta’s Audiobox~\cite{vyas2023audiobox}, Google’s V2A~\cite{deepmind2024generating}, and Veo3~\cite{googleveo32025google} demonstrate promising joint generation of video-synchronized sound and speech.
In the academic community, AudioGen-Omni~\cite{wang2025audiogenomni} struggles with simultaneous sound-speech generation due to limited high-quality paired data. DeepAudio~\cite{zhang2025deepaudio} and DualDub~\cite{tian2025dualdub} integrate speech and sound heads into LLM backbones but rely on complex multi-stage training pipelines for different modules, as end-to-end joint training is believed to degrade performance~\cite{tian2025dualdub}. The feasibility of joint training and generation still remains underexplored in prior works The feasibility of joint training and generation remains largely unexplored in prior works, which our work aims to investigate.

\section{Method}
\label{sec:method}

\subsection{Preliminary}


\paragraph{Flow-Matching Framework}
\label{sec:fm}

ViSSFlow is a flow-matching model built upon Diffusion Transformer (DiT) architecture. Flow matching~\cite{lipman2022flow} learns a velocity field \(v_\theta(x_t, c, t)\) that transforms Gaussian noise \(x_0 \sim \mathcal{N}(0,1)\) into the target audio distribution. The process follows the ODE:
\begin{equation}
\label{eq:fm}
\frac{dx_t}{dt} = v_\theta(x_t, c, t),
\end{equation}
with the training objective:
\begin{equation}
\label{eq:fmloss}
\mathcal{L}_{\text{FM}} = \mathbb{E}_{t, x_0, x_1} \left\| v_\theta(x_t, c, t) - (x_1 - x_0) \right\|^2.
\end{equation}

\paragraph{Diffusion Transformers.}

DiT~\cite{Peebles2022DiT} replaces U-Net backbones with Transformer for generation model. ViSSFlow adopts the same cross-attention-based DiT following Stable Audio Open~\cite{evans2025stable}. In addition to standard cross-attention, we explore concatenating conditioning signals with the latent along the channel dimension, enabling more effective integration of multimodal inputs.

\subsection{ViSSFlow Framework}

\subsubsection{Model Overview.}
As illustrated in Fig.~\ref{fig:arch}, ViSSFlow adopts an attention-based Diffusion Transformer (DiT) architecture inspired by Stable Audio Open~\cite{evans2025stable}. To effectively capture temporal correspondence between audio and video sequences, we apply 1D Rotary Position Embedding (RoPE)~\cite{su2024rope} in both self- and cross-attention layers. The model operates in the latent space of a pretrained audio VAE and reconstructs waveforms using a HiFi-GAN vocoder. 
More details are in Supplementary Material.

\subsubsection{Condition Inputs.}
To enable unified video-conditioned sound and speech generation, ViSSFlow integrates multiple sequence-type conditioning signals. We extract video semantic features using CLIP~\cite{radford2021CLIP} to capture high-level scene and event information. For temporal alignment of environmental sounds, we employ Synchformer features~\cite{iashin2024synchformer}. Speech content is provided through phoneme sequences derived from the input transcript, while AV-HuBERT features~\cite{shi2022avhubert} are used to model lip movements for accurate lip synchronization. All conditioning signals are temporally interpolated to the length of audio latent sequence. Further implementation details are provided in Supplementary Material.

\subsubsection{Disentangled Condition Aggregation Mechanism.} Different conditioning signals exhibit fundamentally different characteristics, and treating them uniformly may lead to suboptimal convergence behaviors and ultimate generation quality.
Within our attention-based DiT architecture, conditioning sequences are incorporated through two main mechanisms:
\textbf{(1) Cross-Attention:} Following the standard implementation of attention-based DiT, conditions are introduced through cross-attention blocks serving as the Key and Value matrices. This mechanism is particularly effective for semantically rich features (e.g., video semantic features $c_v$), where the model needs to globally attend to high-level context for guiding the generation.
\textbf{(2) Channel-wise Concatenation:} Conditions are concatenated with the latent representation $x_t$ along the channel dimension, then processed through the self-attention blocks. This approach has been proven effective in recent V2S~\cite{wang2024frieren} and TTS~\cite{chen-etal-2024-f5tts} frameworks. We find this mechanism more suitable for temporally dense and aligned features (e.g., sound synchronization, speech transcript, and speech synchronization features). By concatenating, the model can more effectively exploit the fine-grained temporal correspondences through self-attention block. 
The final input sequence to the DiT blocks is formulated as: 
\begin{equation}
\begin{split}
x_{in} &= \text{Concat}([x_t, c_{ss}, c_{st}, c_{lip}], \text{dim=channel}) \\
       &\in \mathbb{R}^{T_a \times (D_a + D_{ss} + D_{st} + D_{lip})}.
\end{split}
\end{equation}
Extensive experiments on different condition aggregation mechanisms  and detailed analysis are presented in Sec.~\ref{chap:ab_on_cond}.

\subsection{Training}

\subsubsection{Multimodal Datasets.}
ViSSFlow is trained end-to-end on a combination of sound and speech datasets using the flow-matching loss. The training covers two tasks: (1) Video-to-Sound (V2S), where speech-related conditions are set to zero, and (2) Visual Text-to-Speech (VisualTTS), where sound synchronization features are masked.
For V2S, we use the \textit{VGGSound} dataset~\cite{chen2020vggsound} with 182k training samples. For VisualTTS, we combine three standard benchmarks: \textit{Chem}~\cite{prajwal2020chem} (6,240 training samples), \textit{GRID}~\cite{cooke2006grid} (29,600 training samples), and \textit{LRS2}~\cite{afouras2018LRS2} (126k training samples). In total, the training data consists of approximately 350k samples.
Unlike previous works that rely on complex multi-stage training due to concerns over joint sound-speech optimization~\cite{tian2025dualdub}, ViSSFlow demonstrates strong performance on both tasks through simple end-to-end joint training. More analysis on joint training is provided in Sec.~\ref{chap:ab_on_joint}. Details of the training dataset are provided in Supplementary Material.

\subsubsection{Synthetic Datasets.}

After pretraining, we fine-tune ViSSFlow to enable video-conditioned sound-speech joint generation. To address the scarcity of joint data, we propose an efficient data synthesis strategy that constructs joint video-sound-speech samples in the feature space. We randomly select sound and speech samples from \textit{VGGSound} and \textit{LRS2}, respectively. Given a sound waveform $w_a$ and a speech waveform $w_s$, we define a temporal shift operator $S_{\tau}(\cdot)$ that pads speech segment to a random start time $\tau$. We implement two synthesis modes to simulate diverse acoustic scenarios:
\begin{itemize}
\item \textbf{Additive Synthesis (Add):} Speech $(w_s)$ is superimposed onto the background sound $(w_a)$ to simulate overlapping audio $(w^{syn})$ at a random Signal-to-Noise Ratio $\alpha$ :
\begin{equation}
\label{eq:add}
w^{syn} = w_a + \alpha \cdot S_{\tau}(w_s), \quad c_{v}^{syn} = c_{v}^{a} + \alpha \cdot S_{\tau}(c_{v}^{s})
\end{equation}

\item \textbf{In-Place Substitution (Insert):} The background content of $w_a$ within the speech interval is replaced by $w_s$, simulating  $w^{syn}$ with clean transition between modalities:
\begin{equation}
\begin{split}
\label{eq:insert}
    w^{syn} &= (1 - \mathbf{m}_{\tau}) \odot w_a + S_{\tau}(w_s), \\
    c_{v}^{syn} &= (1 - \mathbf{m}_{\tau}) \odot c_{v}^{a} + S_{\tau}(c_{v}^{s})
\end{split}
\end{equation}
where $\mathbf{m}_{\tau}$ is a binary mask corresponding to the speech interval of $S_{\tau}$.
\end{itemize}

During synthesis, we retain the sound synchronization features from the sound source, as well as the speech transcript and speech synchronization features from the speech source. 
The video semantic features of sound and speech ($c_{v}^{a}, c_{v}^{s}$) are implemented the same operation as the wave data to form the synthetic video feature ($c_{v}^{syn}$), as in Eqs.~\ref{eq:add} and \ref{eq:insert}. Since these operations are performed in the feature space during data loading, they bypass the need to modify raw video and audio data.
Consequently, this synthesis method introduces negligible computational overhead while accommodating a wide range of joint scenarios. We synthesize over 2.8M synthetic samples on the fly.
By fine-tuning with easily accessible synthetic data, ViSSFlow can be adapted to joint sound-speech generation. This demonstrates the robust capabilities of ViSSFlow and validates the efficacy of our data synthesis approach. Detailed analysis is provided in Sec.~\ref{chap:ab_on_synth}.

\subsubsection{Implementation Detail.}

During training, audio samples are padded or truncated to 10 seconds and resampled to 16 kHz. We use a learning rate of $4 \times 10^{-7}$ with 2,000 warm-up steps. The model is trained on 8 NVIDIA H100 GPUs with a batch size of 36 per GPU. The unconditional probability is set to 0.1 to support classifier-free guidance (CFG).
We train three variants of ViSSFlow with different sizes: ViSSFlow-S (5 DiT blocks, 245M parameters), ViSSFlow-M (10 DiT blocks, 463M parameters), and ViSSFlow-L (15 DiT blocks, 618M parameters). All parameters of the DiT backbone are fully optimized, while the VAE and vocoder remain frozen. The joint multi-task training phase is conducted for 250k steps. Subsequently, the fine-tuning phase for joint generation is performed for 10k steps based on the pretrained ViSSFlow using synthetic data only. 
During inference, we employ the Dopri5 solver to perform 50-step ODE integration for denoising. A CFG scale of 3.0 is applied across all tasks. 
Unless otherwise specified, all results reported in the main paper are obtained using ViSSFlow-L. Comprehensive evaluation results for all model variants, along with additional implementation details and ablation studies on the CFG scale, are provided in Supplementary Material.

\section{Experiment}
\label{sec:exp}
\begin{table}[t]
\centering
\setlength{\tabcolsep}{1.6pt}
\begin{tabular}{l|cccc|c|c}
\toprule[1.2pt]

\multicolumn{1}{c|}{} &
\multicolumn{4}{c|}{\textit{Sound Quality}} &
\multicolumn{1}{c|}{\textit{Synch.}} &
\textit{Seman.} \\

\midrule

\multirow{2}{*}{Method}
& $\mathrm{FAD}\downarrow$
& $\mathrm{FAD}\downarrow$
& $\mathrm{KL}\downarrow$
& $\mathrm{KL}\downarrow$
& \multicolumn{1}{c}{DeSync}
& \multicolumn{1}{c}{IB-VA} \\

& (pas.)
& (pann.)
& (pas.)
& \multicolumn{1}{c|}{(pann.)}
& \multicolumn{1}{c|}{$\downarrow$}
& $\uparrow$ \\

\midrule[1.2pt]

SpecVQ. & 270.07 & 28.62 & 3.20 & 3.31 & 1.23 & 13.77 \\
Im2Wav  & 242.05 & 18.34 & 2.27 & 2.50 & 1.22 & 19.59 \\
V-AURA  & 218.50 & 14.80 & 2.76 & 2.42 & \underline{0.65} & 27.60 \\

\midrule

VAB & 192.07 & 18.26 & 2.36 & 2.51 & 1.17 & 25.82 \\
\color{gray!40}MultiF.
& \color{gray!40}148.93
& \color{gray!40}13.81
& \color{gray!40}1.51
& \color{gray!40}1.69
& \color{gray!40}0.82
& \color{gray!40}27.00 \\

\midrule

Difffoley & 409.20 & 23.05 & 3.09 & 3.14 & 0.91 & 20.60 \\
Seeing.   & 206.63 & 21.96 & 2.66 & 2.78 & 1.20 & \color{gray!40}36.81 \\
V2A-M.    & 113.55 & \underline{8.16} & 2.47 & 2.66 & 1.22 & 22.25 \\
FoleyC.   & 142.34 & 12.85 & 2.33 & 2.33 & 1.21 & \underline{27.75} \\
TiVA      & 304.60 & 23.62 & 2.78 & 3.23 & 1.18 & 16.92 \\
LoVA      & 150.54 & 13.13 & \underline{2.14} & \underline{2.30} & 1.22 & 26.01 \\

\midrule

Frieren & \underline{100.33} & 11.66 & 2.80 & 2.71 & 0.85 & 22.88 \\

\color{gray!40}MMAudio
& \color{gray!40}65.16
& \color{gray!40}5.95
& \color{gray!40}1.42
& \color{gray!40}1.67
& \color{gray!40}0.45
& \color{gray!40}33.23 \\

\rowcolor{gray!20}
ViSSFlow
& \textbf{98.45}
& \textbf{6.52}
& \textbf{2.08}
& \textbf{2.26}
& \textbf{0.59}
& \textbf{29.59} \\

\bottomrule[1.2pt]
\end{tabular}

\caption{
V2S evaluation results on the VGGSound benchmark. 
For each metric, the highest score is shown in \textbf{bold}, and the second-highest is \underline{underlined}.
}
\label{tab:v2a}
\end{table}

\begin{table}[t]
\centering
\setlength{\tabcolsep}{1.2pt}

\begin{tabular}{l|cccc|cc}
\toprule[1.2pt]

&
\multicolumn{4}{c|}{\textit{Speech Quality}} &
\multicolumn{2}{c}{\textit{Synch.}} \\
\midrule

\multirow{2}{*}{{Method}}
& {WER}
& {Spk.}
& {UTMOS}
& {MCD}
& {LSE-C}
& {LSE-D} \\

& $\downarrow$
& Sim. $\uparrow$
& $\uparrow$
& $\downarrow$
& $\uparrow$
& $\downarrow$ \\

\toprule[1.2pt]
\multicolumn{7}{l}{\textit{Benchmark -- Chem}}\\
\midrule



DSU & 37.3 & 72.9 & \underline{3.33} & \underline{10.52} & 5.88 & 7.86 \\
HPMDubbing & 22.3 & 44.6 & 3.11 & 11.54 & \underline{7.30} & \underline{7.58} \\
StyleDubber & \underline{16.3} & 73.7 & 3.14 & 14.46 & 3.58 & 11.08 \\
EmoDubber & 16.7 & \textbf{78.1} & \textbf{3.87} & 14.87 & 5.48 & 8.27 \\

\midrule

\rowcolor{gray!20}
ViSSFlow & \textbf{9.2} & \underline{76.1} & 3.31 & \textbf{7.96} & \textbf{7.89} & \textbf{6.73} \\

\toprule[1.2pt]
\multicolumn{7}{l}{\textit{Benchmark -- GRID}}\\
\midrule



DSU & 34.3 & 5.9 & 3.55 & 13.82 & 5.63 & \underline{8.73} \\
HPMDubbing & 27.6 & 31.3 & 2.11 & \underline{12.31} & 6.02 & 8.85 \\
StyleDubber & \textbf{10.9} & \textbf{51.4} & \underline{3.74} & 12.85 & \underline{6.33} & 8.77 \\
EmoDubber & \underline{15.9} & 50.5 & \textbf{3.98} & 15.52 & 3.48 & 10.35 \\

\midrule

\rowcolor{gray!20}
ViSSFlow & {16.2} & \textbf{51.4} & 3.30 & \textbf{8.62} & \textbf{6.81} & \textbf{8.18} \\

\bottomrule[1.2pt]
\end{tabular}

\caption{
VisualTTS evaluation results on the Chem and GRID benchmarks. 
}
\label{tab:VisualTTS}

\end{table}

\begin{table*}[t!]
    \centering
    \setlength{\tabcolsep}{2pt}
    \begin{tabular}{l|cccccc|cccc}
        \toprule[1.2pt]
         
        & \multicolumn{6}{c|}{Speech} 
        & \multicolumn{4}{c}{Audio} \\
        \midrule

        Method
        & WER $\downarrow$ 
        & Spk. Sim. $\uparrow$ 
        & UTMOS $\uparrow$
        & MCD $\downarrow$ 
        & LSE-C $\uparrow$
        & LSE-D $\downarrow$
        & FAD$_{\text{pas.}}$ $\downarrow$ 
        & FAD$_{\text{pann.}}$ $\downarrow$ 
        & KL$_{\text{pas.}}$ $\downarrow$ 
        & KL$_{\text{pann.}}$ $\downarrow$ \\
        \midrule[1.2pt]
        
        LoVA+Speaker
        & \underline{25.98} & \underline{20.02} & \underline{1.33} & 23.77 & 1.28 & \underline{10.66}
        & 285.84 & 8.68 & 1.27 & 0.97 \\
        LoVA+Style
        & 37.56 & 19.58 & 1.30 & \underline{23.05} & 1.27 & \textbf{10.65}
        & 296.64 & 9.28 & 1.31 & 1.05 \\
        MMAudio+Speaker
        & 41.47 & 18.39 & 1.32 & 24.15 & 1.91 & 10.95
        & 229.65 & \textbf{4.32} & \underline{1.14} & \textbf{0.89} \\
        MMAudio+Style
        & 56.86 & 17.78 & 1.30 & 23.33 & \underline{1.92} & 10.86
        & 240.70 & \underline{4.48} & 1.16 & \underline{0.91} \\
        \midrule
        DeepAudio
        & 39.20 & 12.30 & 1.31 & 27.76 & 1.85 & 10.98
        & \underline{218.01} & 8.72 & 1.43 & 1.35 \\
        \rowcolor{gray!20}
        ViSSFlow
        & \textbf{19.40} & \textbf{24.70} & \textbf{2.07} & \textbf{17.36} & \textbf{2.00} & 11.43
        & \textbf{177.16} & 6.83 & \textbf{1.07} & \underline{0.91} \\
        \bottomrule[1.2pt]
    \end{tabular}
    \caption{Video-conditioned sound-speech joint generation evaluation results on the V2C benchmark.}
    \label{tab:v2c_full}
\end{table*}

\subsection{Experiment Settings}




\subsubsection{Benchmarks.}
For V2S, we report results on the VGGSound~\cite{chen2020vggsound} test set (15k samples). For VisualTTS, we use the test sets from Chem (200 samples)~\cite{prajwal2020chem}, GRID (3,291 samples)~\cite{cooke2006grid}, and LRS2 (700 samples)~\cite{afouras2018LRS2}. We conduct evaluation of video-to-sound-speech joint generation on the V2C-Animation~\cite{chen2022v2c} test set following the evaluation protocol of DualDub~\cite{tian2025dualdub}. This dataset consists of 1,300 video segments featuring synchronized speech and environmental sounds with high temporal alignment. 
More detailed evaluation settings and baseline configurations can be found in Supplementary Material.

\subsubsection{Metrics.} For V2S, we evaluate sound quality using Fréchet Audio Distance (FAD)~\cite{kilgour2018frechet} and Mean KL-Divergence (KL) following MMAudio~\cite{cheng2025MMAudio}. Features are extracted by the audio classifiers PaSST~\cite{koutini22passt} and PANN~\cite{kong2020panns}. Temporal alignment is measured by the DeSync Score, computed via SynchFormer features~\cite{iashin2024synchformer}. Sound-visual relevance is assessed by cosine similarity between video and audio ImageBind embeddings(VA-IB)~\cite{girdhar2023imagebind}.
For VisualTTS, we use Word Error Rate (WER) for intelligibility and UTokyo-SaruLab Mean Opinion Score (UTMOS)~\cite{saeki2022utmos} for perceptual  quality. 
Speaker style similarity is measured with Versa~\cite{shi2024versa}. Synchronization is quantified by Mel-Cepstral Distortion (MCD) via the ESPNet toolkit~\cite{watanabe2018espnet} with Dynamic Time Warping. Lip synchronization is assessed using Lip Sync Error Distance (LSE-D) and Lip Sync Error Confidence (LSE-C) computed by SyncNet~\cite{Chung16syncnet}.


\subsection{Automatic Evaluation on Benchmarks}

\subsubsection{V2S and VisualTTS Benchmarks.}
Tab.~\ref{tab:v2a} presents the performance of ViSSFlow on the VGGSound test set. To ensure a comprehensive evaluation, we compare against representative autoregressive, mask, diffusion, and flow-based V2S models. 
For fairness, models trained with additional high-quality Text-to-Audio datasets (e.g., MMAudio and MultiFoley) are marked in gray and not directly compared~\cite{cheng2025MMAudio}.
As shown in the table, ViSSFlow achieves state-of-the-art results across audio quality, semantic alignment, and temporal synchronization. Notably, even though ViSSFlow is a unified model rather than V2S-specialized, it still delivers superior overall performance. 

Tab.~\ref{tab:VisualTTS} summarizes ViSSFlow’s performance on the Chem and GRID benchmarks. We compare it with representative VisualTTS methods including DSU~\cite{lu23fDSU}, HPMDubbing~\cite{cong2023hpmdubbing}, StyleDubber~\cite{cong2024styledubber}, and EmoDubber~\cite{cong2024emodubber}.
As shown in the table, ViSSFlow consistently outperforms prior methods in speech quality (WER, MCD) and speaker similarity (Spk. Sim.). 
Moreover, it establishes new state-of-the-art results on synchronization metrics (LSE-C, and LSE-D) on both benchmarks. These results highlight the effectiveness of our disentangled condition aggregation mechanism in generating high-quality speech with precise lip synchronization. 
However, a limitation is that the UTMOS score is constrained by the upper bound of the VAE-Vocoder pipeline. More analysis is provided in Supplementary Material.


\subsubsection{Joint Generation Benchmark.}

Tab.~\ref{tab:v2c_full} compares ViSSFlow with the recent unified framework DeepAudio~\cite{zhang2025deepaudio}, as well as several competitive pipeline-based baselines following~\cite{tian2025dualdub}. Specifically, the pipeline methods independently generate speech using VisualTTS models and environmental sounds using V2S models, then merge the two outputs into the final audio track.
ViSSFlow consistently outperforms both the unified baseline and the pipeline-based approaches across most speech and sound evaluation metrics. Notably, all speech generation baselines (DeepAudio, Speaker2Dubber, and StyleDubber) are trained on the in-domain V2C data samples, whereas ViSSFlow is adapted using only out-of-domain synthetic data. The pretraining stage and the condition mechanism equip ViSSFlow with a strong unified audio generation prior, enabling effective transfer to the joint sound-and-speech distribution with lightweight joint data. These results demonstrate the advantage of our unified generation framework as well as the effectiveness of the synthetic data generation strategy.

\subsection{Human Evaluation of Joint Generation}
We conduct a human study on 15 randomly selected V2C-Animation videos and 15 OOD Veo3-generated videos. 
Ten raters evaluate all methods using a  blind Mean Opinion Score protocol (score from 1 to 5) across five aspects: speech quality, speech synchronization, speaker similarity, sound quality, and sound synchronization. As shown in Tab.~\ref{tab:human_eval_combined}, ViSSFlow consistently outperforms existing open-source baselines across all metrics, demonstrating its superior generalizability and capability in realistic joint generation.

\begin{table}[h!]
    \centering
    \setlength{\tabcolsep}{1.8pt}

    \begin{tabular}{l|ccc|cc}
        \toprule[1.2pt]

        & \multicolumn{3}{c|}{Speech}
        & \multicolumn{2}{c}{Sound} \\
        \midrule

        Method
        & Qual.$\uparrow$
        & Sync.$\uparrow$
        & Spk. Sim.$\uparrow$
        & Qual.$\uparrow$
        & Sync.$\uparrow$ \\
        \midrule[1.2pt]

        LoVA+Speaker
        & 2.65 & 2.57 & 2.63
        & 2.59 & 2.43 \\

        MMAudio+Style
        & \underline{2.91} & \underline{2.96} & 2.89
        & {2.88} & {2.69} \\

        DeepAudio
        & 2.90 & 2.92 & \underline{2.95}
        & \underline{3.05} & \underline{2.84} \\

        
        \midrule
        
        \rowcolor{gray!20}
        ViSSFlow (Ours)
        & \textbf{3.57} & \textbf{3.23} & \textbf{3.44}
        & \textbf{3.48} & \textbf{3.30} \\
        
        \bottomrule[1.2pt]
    \end{tabular}

    \caption{
    Human evaluation on 30 joint generation samples from V2C and out-of-domain Veo3-generated videos. 
    }

    \label{tab:human_eval_combined}
\end{table}



\begin{figure*}[h]
    \centering
    \includegraphics[width=0.98\linewidth]{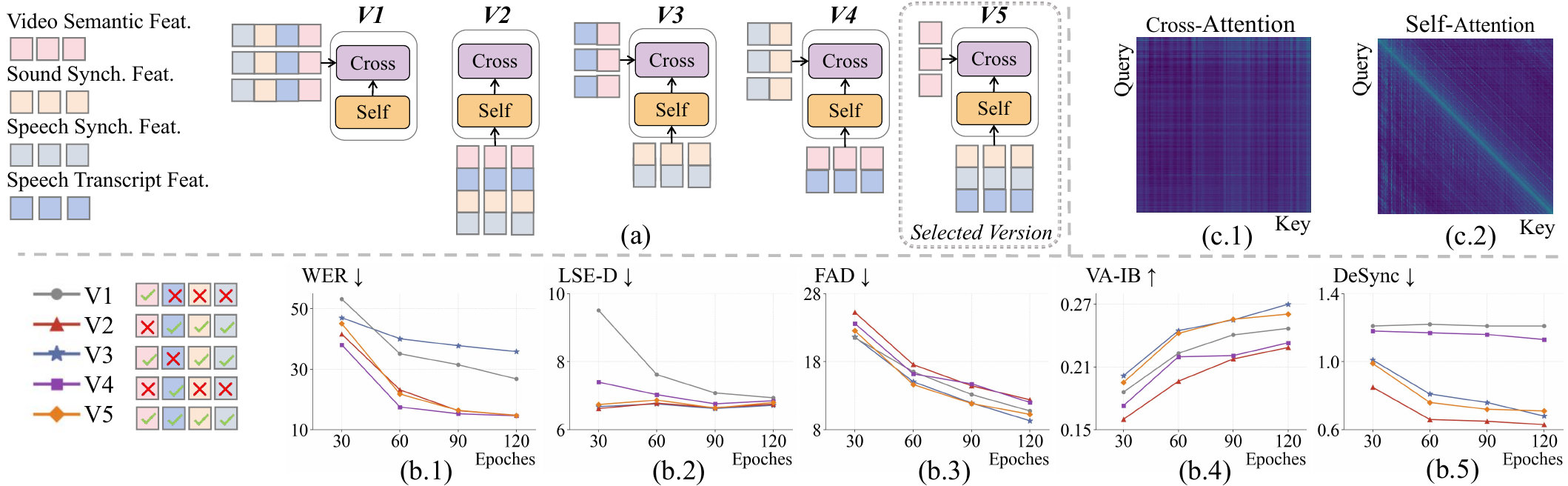}
    \caption{
    {
    Performance comparison of different conditioning mechanisms.
    (a) demonstrates the overview of different model architectures.
    (b.1) and (b.2) show WER and LSE-D metrics for the VisualTTS task, while (b.3) - (b.5) present FAD, VA-IB, and DeSync metrics for the V2S task. 
    The checkmarks and crosses in the legend of (b) indicate whether the variant effectively incorporates the specific condition.
    (c) is the visualization of the attention weights in self- and cross-attention layers of DiT.
    }
    }
    \label{fig:ablation_arch}
\end{figure*}

\subsection{Ablation of Condition Mechanism Design}
\label{chap:ab_on_cond}

To demonstrate the advantage of our disentangled  condition aggregation mechanism for integrating multimodal signals, we investigate five variants as illustrated in Fig.~\ref{fig:ablation_arch}(a). These variants differ in how they incorporate video semantic features $c_{v}$, sound synchronization features $c_{ss}$, speech transcript features $c_{st}$, and speech synchronization features $c_{lip}$. 
We trained each variant for 120 epochs on the combined V2S and VisualTTS datasets. The convergence curves for speech and sound metrics are shown in Fig.~\ref{fig:ablation_arch}(b). From these results, we derive the following observations: 
\textbf{(1) Features with explicit temporal alignment properties are best suited for integration via concatenation.}
Specifically, variants incorporating speech transcript $c_{st}$ via concatenation (V2, V4, V5) consistently outperform others (V1, V3) in WER metrics as in Fig.~\ref{fig:ablation_arch}(b.1). 
Similarly, for sound synchronization features ($c_{ss}$) and speech synchronization features ($c_{lip}$), variants V2, V3, and V5 exhibit superior DeSync and LSE-D compared to V1 and V4, demonstrated in Fig.~\ref{fig:ablation_arch}(b.2) and (b.5). As training progresses, the gap in LSE-D gradually diminishes. However, the discrepancy in DeSync remains relatively persistent.
We hypothesize that conditions introduced via concatenation are primarily processed through the model's self-attention layers. Our visualization of ViSSFlow's self-attention heatmap in Fig.~\ref{fig:ablation_arch}(c.2) reveals a prominent diagonal inductive bias, indicating that the self-attention layers primarily focus on local context from neighboring positions. This finding is consistent with prior literature~\cite{li2023selfattn, liu2024selfattn}. For information with highly local correlations, the self-attention layers effectively leverage this diagonal bias to accelerate convergence and achieve finer temporal granularity.
\textbf{(2) Global semantic information is more effectively integrated through cross-attention.} Our results indicate that V1, V3, and V5 consistently achieve higher performance in sound generation metrics (FAD in Fig.~\ref{fig:ablation_arch}(b.2) and VA-IB in Fig.~\ref{fig:ablation_arch}(b.4)). For features that contain relatively sparse temporal information like video semantic embeddings, forced concatenation may hinder the learning of semantic correlations due to the restrictive diagonal bias of self-attention. Instead, cross-attention provides a more flexible mechanism for free-form interaction (see Fig.~\ref{fig:ablation_arch}(c.1)), allowing the model to dynamically attend to global visual cues without being constrained by rigid temporal structures.
Based on these insights, we do not treat all features uniformly and instead derive the \textbf{disentangled condition aggregation mechanism} as in Fig.~\ref{fig:arch}. By concatenating temporally sensitive features ($c_{ss}, c_{lip}, c_{st}$) and utilizing cross-attention for global semantic features ($c_{v}$), ViSSFlow strikes an optimal performance on both temporal synchronization and semantic fidelity.

\subsection{Other Ablations}

\begin{figure}[t!]
    \centering
    \includegraphics[width=1.0\linewidth]{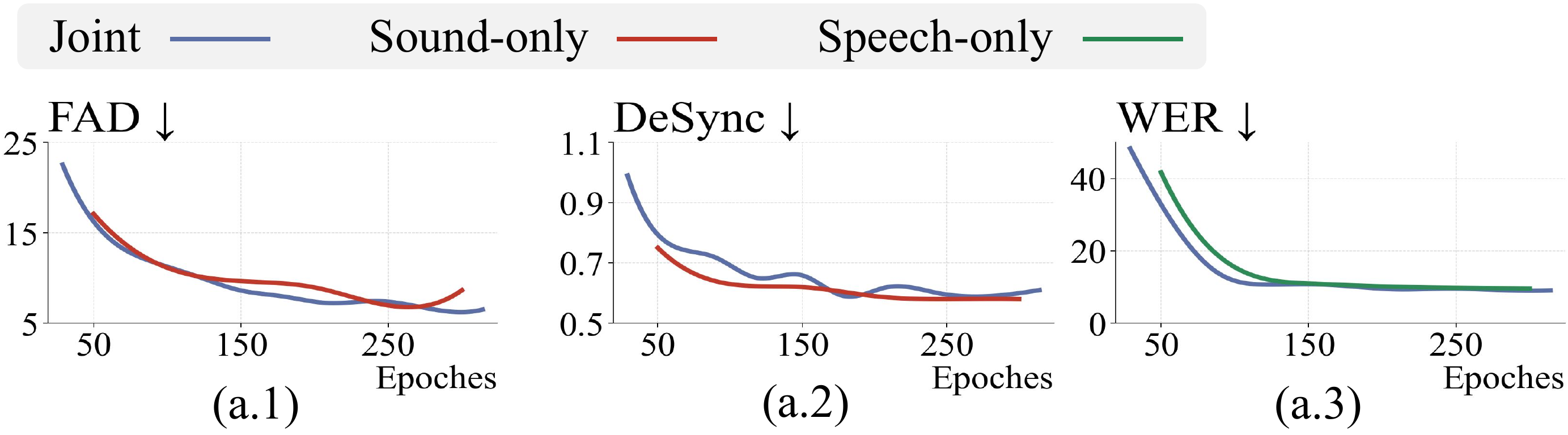}
    \caption{
        Impact of joint learning on sound and speech generation.
        Panels (a.1), (a.2) display FAD and DeSync metrics for the V2S task over training epochs, 
        while (a.3) presents WER for VisualTTS. The joint learning does not compromise convergence speed or final synthesis quality.
    }
    \label{fig:ablation_data}
\end{figure}

\subsubsection{Comparison of Joint and Single-Task Training.}
\label{chap:ab_on_joint}

We investigate the impact of joint training with sound and speech data. We compare ViSSFlow under three distinct data configurations: sound data-only, speech data-only, and joint training (sound and speech data). The learning curves illustrated in Fig.~\ref{fig:ablation_data} provide a direct visualization of the convergence behavior across these settings. 
As observed, at the same training epoch, the joint training configuration achieves performance identical to that trained with the speech-only or sound-only data. This suggests that joint learning does not lead to mutual suppression or suboptimal convergence. Instead, the two modalities can be learned concurrently within a unified framework without compromising the generative quality of either domain. 
Therefore, contrary to the previous belief~\cite{tian2025dualdub}, the learning processes of ViSSFlow do not require additional designs on training stages.

\subsubsection{Effectiveness of Synthetic Data.}
\label{chap:ab_on_synth}
To investigate the efficacy of our data synthesis method, we conduct an ablation study on the V2C benchmark with three settings: V2C training data only (8k training samples), synthesized data only, and V2C+synthesized data. All variants are fine-tuned for 10k steps.
As shown in Tab.~\ref{tab:data_ablation}, synthetic data plays a critical role in joint generation. Training solely on V2C leads to high WER, low UTMOS, and poor synchronization due to limited scale and domain diversity. 
In contrast, incorporating synthesized data significantly improves speech intelligibility, lip synchronization, and sound quality. 
Notably, the model trained with synthesized data alone achieves the best WER and UTMOS. Furthermore, synthetic data greatly enhances generalization, as demonstrated by strong zero-shot performance on out-of-domain Veo3-generated videos (see Fig.~\ref{fig:teaser}(c) and Tab.~\ref{tab:human_eval_combined}). 
These results highlight the importance of high-quality synthetic data for helping unified models adapt to complex joint sound-speech distributions. More demos are available in Supplementary Material.

\begin{table}[h]
    \centering
    \setlength{\tabcolsep}{1.5pt}
    \begin{tabular}{l|ccc|cc}
        \toprule[1.2pt]

        
        
        \multirow{1}{*}{Training Data} & WER$\downarrow$ & UTMOS$\uparrow$ & LSE-C$\uparrow$
        & FAD$\downarrow$ & KL$\downarrow$ \\
        \midrule[1.2pt]

        V2C
        & 36.5
        & 1.45
        & \underline{2.42}
        & \underline{158.01}
        & \underline{1.01} \\

        V2C + Synth. Data
        & \underline{22.6}
        & \underline{1.76}
        & \textbf{2.58}
        & \textbf{116.08}
        & \textbf{1.00} \\

        Synthesized Data
        & \textbf{19.4}
        & \textbf{2.07}
        & 2.00
        & 177.16
        & 1.07 \\

        \bottomrule[1.2pt]
    \end{tabular}

    \caption{
    Experiment on training data configurations for video-conditioned audio-speech joint generation on the V2C benchmark. The introduction of synthetic data significantly improves both sound and speech generation quality.
    }
    \label{tab:data_ablation}

\end{table}

\paragraph{Ablation on Multi-modal Conditions.} 
We evaluate the contribution of each conditioning signal by randomly dropping individual features during training and applying the same dropout at inference. 
For sound synchronization features, we additionally compare SynchFormer with CAVP from difffloey~\cite{luo2023difffoley}. As shown in Tab.~\ref{tab:ablation_conditions}, each condition plays a distinct role: removing CLIP significantly degrades sound quality (FAD), while dropping or replacing synchronization features substantially harms temporal alignment (DeSync). Transcript features are critical for speech intelligibility (WER), and lip features extracted by AV-HuBERT significantly improve lip synchronization (LSE-C). Overall, the selected conditioning signals provide complementary benefits and yield the best performance.

\begin{table}[h!]
    \centering
    \setlength{\tabcolsep}{1.5pt}
    \begin{tabular}{l|cccc}
        \toprule[1.2pt]

        \multirow{1}{*}{Condition}
        & WER$\downarrow$
        & LSE-C$\uparrow$
        & FAD$\downarrow$
        & DeSync$\downarrow$ \\
        \midrule[1.2pt]

        \rowcolor{gray!20}
        Full
        & 14.2
        & 8.79
        & 6.68
        & 0.58 \\

        w/o CLIP
        & 14.1 {\scriptsize\textcolor{gray}{(-0.1)}}
        & 9.16 {\scriptsize\textcolor{gray}{(+0.37)}}
        & 20.42 {\scriptsize\textcolor{black}{\textbf{(+13.74)}}}
        & 0.63 {\scriptsize\textcolor{gray}{(+0.05)}} \\

        w/o Synch
        & 14.8 {\scriptsize\textcolor{gray}{(+0.6)}}
        & 8.77 {\scriptsize\textcolor{gray}{(-0.02)}}
        & 10.57 {\scriptsize\textcolor{gray}{(+3.89)}}
        & 1.24 {\scriptsize\textcolor{black}{\textbf{(+0.66)}}} \\

        w/o Text
        & 54.4 {\scriptsize\textcolor{black}{\textbf{(+40.2)}}}
        & 8.88 {\scriptsize\textcolor{gray}{(+0.09)}}
        & 6.68 {\scriptsize\textcolor{gray}{(--)}}
        & 0.58 {\scriptsize\textcolor{gray}{(--)}} \\

        w/o Lip
        & 15.7 {\scriptsize\textcolor{gray}{(+1.5)}}
        & 8.02 {\scriptsize\textcolor{black}{\textbf{(-0.77)}}}
        & 6.68 {\scriptsize\textcolor{gray}{(--)}}
        & 0.58 {\scriptsize\textcolor{gray}{(--)}} \\

        w/ CAVP
        & 14.7 {\scriptsize\textcolor{gray}{(+0.5)}}
        & 8.72 {\scriptsize\textcolor{gray}{(-0.07)}}
        & 11.47 {\scriptsize\textcolor{gray}{(+4.79)}}
        & 0.99 {\scriptsize\textcolor{black}{\textbf{(+0.41)}}} \\

        \bottomrule[1.2pt]
    \end{tabular}

    \caption{
    Ablation on different conditions.
    Values in parentheses denote the performance change relative to the full model.
    }

    \label{tab:ablation_conditions}
\end{table}

\section{Conclusions}
\label{sec:conc}


This work introduces ViSSFlow, a unified flow-matching framework that enables effective joint video-conditioned sound and speech generation through disentangled condition aggregation. With the data synthesis strategy, we alleviate data scarcity and improve joint generation performance. For future work, developing more compact representations that better preserve speech details remains an important direction.


\bibliography{aaai2027}


\end{document}